\documentstyle[prc,aps,psfig,twocolumn]{revtex}
\begin{document}
\draft
\title{$^{180}$Ta production in the classical s-process}
\author{Markus Loewe$^1$, Petr Alexa$^2$, Jorrit de Boer$^1$, 
Michael W\"urkner$^1$}
\address{$^1$Sektion Physik, Munich University, Am Coulombwall 1, Garching, 
D-85768, Germany}
\address{$^2$Department of Physics and Measurements, Institute of
Chemical Technology, \\ 
Technick\' a 5, CZ-166 28 Prague 6, Czech Republic}
\date{\today}
\maketitle

\begin{abstract}
The production and survival of the quasistable isomer $^{180}$Ta during the 
stellar nucleosynthesis has remained a matter of discussion for years. 
A careful analysis 
of the available experimental data and theoretical calculations enabled us 
to reproduce the observed solar abundance of $^{180}$Ta even in the classical 
s-process ($kT=28$ keV -- 33 keV). 
\end{abstract}


\vspace*{-0.5 cm}
$^{180}$Ta is the only nucleus present in nature in an $9^-$ isomeric state 
($^{180}$Ta$^{m}$) at an energy of 75.3 keV. For its 
half-life an experimental lower limit of $1.2$ $\times $ 10$^{15}$ y 
was obtained \cite{Cummings}. The $1^+$ ground state of $^{180}$Ta 
decays to $^{180}$Hf and $^{180}$W (half-life 8.15 h). 
$^{180}$Ta$^{m}$ has a very small abundance: Only $0.012\%$ of 
natural Ta consists of $^{180}$Ta$^{m}$, the rest is $^{181}$Ta. The
solar abundance of $^{180}$Ta$^{m}$ (normalized to Si with the
solar abundance of $10^{12}$) is 2.48 (cf.~solar abundances of
neighbouring isotopes: $4.20 \cdot 10^4$ for $^{178}$Hf, $2.10 \cdot 10^4$ for
$^{179}$Hf, $5.4 \cdot 10^4$ for $^{180}$Hf, and $173$ for $^{180}$W) 
\cite{Anders}.
     
To explain the production of $^{180}$Ta$^{m}$ during the nucleosynthesis,
two types of production processes were proposed. 

\medskip

\noindent
I. Explosive processes occur in supernovae:
\begin{enumerate}
\vspace*{-0.2 cm}
\item The r-process leading to an isomer in $^{180}$Lu decaying to 
an $8^-$ isomer in $^{180}$Hf at $1.14$ MeV, further decaying partially to 
$^{180}$Ta$^{m}$ \cite{eschner:84,kellogg:92}:
The amount of $^{180}$Ta$^{m}$ produced in the r-process relative to the
amount of $^{180}$Hf can be obtained from~\cite{beer:82s}:
\begin{equation}  
\frac{N_r(^{180}\mbox{Ta}^{m})}{N_r(^{180}\mbox{Hf})}=f^{180}_{m}\cdot 
f^{m}_{\beta ^-} \ ,
\end{equation}
where $f^{180}_{m}$ and $f^{m}_{\beta ^-}$ are the branching factors for
the $\beta ^-$ or $\beta ^- + \gamma $ transitions $^{180}$Lu $\rightarrow $
$^{180}$Hf$^{m}$ and $^{180}$Hf$^{m}$ $\rightarrow $ $^{180}$Ta$^{m}$, 
respectively. For $f^{m}_{\beta ^-}$ a value of $(0.29\pm0.08)\%$ was
found \cite{kellogg:92}. The laboratory value can also be used for the 
r-process. Corrections due to the high degree of
ionization of $^{180}$Hf$^{m}$ can be neglected since the r-process duration
($\le 100$ s) is short compared to the isomer half-life (5.5 hr) and
the temperature and density are greatly diminished when the decay of
$^{180}$Hf$^{m}$ occurs. Values of $(0.46\pm0.15)\%$ \cite{eschner:84} or 
$(0.005\pm0.018)\%$ \cite{kellogg:86} were found for $f^{180}_{m}$. 
These two values of $f^{180}_{m}$ lead to ($9\pm4$)\% of the observed 
$^{180}$Ta$^{m}$ solar abundance or to a negligible $^{180}$Ta$^{m}$
production. 
\vspace*{-0.2 cm}
\item In the rapid p-process, $^{180}$Ta$^{m}$ can be produced in the
$^{181}$Ta($\gamma$,n) reaction (a negligible production
\cite{woosley:78} or an overproduction comparable to that of neighboring
nuclei \cite{rayet:95}). 
\vspace*{-0.2 cm}
\item In the $\nu$-process during the supernova-core collapse into a 
neutron star, $^{180}$Ta$^{m}$ can be produced in the 
$^{181}$Ta($\nu$,$\nu '$n) reaction \cite{woosley:90,Belnew}.   
\end{enumerate}

\noindent
II. Non-explosive processes can be summarized as follows:
\begin{enumerate}
\vspace*{-0.2 cm}
\item The s-process branching in $^{180}$Hf: In the s-process via 
$^{179}$Hf(n,$\gamma$), the $8^-$ isomer in $^{180}$Hf partially decaying to
$^{180}$Ta$^{m}$ (with the branching factor $f^{m}_{\beta ^-}$)
is populated with the branching factor $^{\mbox{\small{Hf}}}\!f^m_{n}$ or 
$B$ \cite{warde81}.   
We assume that the classical s-process lasts more than 1 year, the 
temperature lies 
between $kT=28$ keV and 33 keV \cite{Wisshak:95}, the neutron density 
$n_n = (4.1\pm 0.6)\cdot 10^8~\mbox{cm}^{-3}$ \cite{Toukan:95}, the electron 
density $n_e = 5.4\cdot 10^{26}~\mbox{cm}^{-3}$ \cite{Best:96}, and $kT$, 
$n_n$, and $n_e$ remain constant during the s-process.
For the relative abundance of $^{180}$Ta$^{m}$ one then obtains:
\begin{equation}\label{s-prozessformelnochmal180Hf}
\left[\frac{N_s(^{180}\mbox{Ta}^{m})}{N_s(^{179}\mbox{Hf})}\right]_1=
\frac{\langle \sigma \rangle_{^{179}\mbox{\small{Hf+n}}}}{\langle \sigma 
\rangle_{^{180}\mbox{\small{Ta}}^{m}\mbox{\small{+n}}}}\cdot 
^{\mbox{\small{Hf}}}\!\!f^m_{n} \cdot f^{m}_{\beta ^-} \ ,
\end{equation}
where $\langle \sigma \rangle_{^{179}\mbox{\small{Hf+n}}}$ and $\langle \sigma 
\rangle_{^{180}\mbox{\small{Ta}}^{m}\mbox{\small{+n}}}$ are Maxwellian 
averaged neutron-capture cross sections taken from \cite{beer:82s} 
(Table VIII) and \cite{Wisshak:00,Wisshak} (Table 9), respectively.
For $\langle \sigma \rangle_{^{179}\mbox{\small{Hf+n}}} = (991\pm 30)$ mb, 
$\langle \sigma \rangle_{^{180}\mbox{\small{Ta}}^{m}\mbox{\small{+n}}} = 
(1465\pm 100)$ mb ($kT = 30$ keV), 
$^{\mbox{\small{Hf}}}\!f^m_{n} = 
\sigma^m(^{179}\mbox{Hf})/\sigma(^{179}\mbox{Hf}) =(1.24\pm0.06)\%$ 
\cite{beer:82s}, and $f^{m}_{\beta ^-} \approx 0.7\%$ \cite{beer:82s},
this s-process branching can account for only ($16\pm3$)\% of the observed
$^{180}$Ta$^{m}$ solar abundance.
\vspace*{-0.2 cm}
\item The s-process branching in $^{179}$Hf: Excited states
in $^{179}$Hf decaying to $^{179}$Ta can be thermally populated. 
$^{180}$Ta$^{m}$ is then produced in the $^{179}$Ta(n,$\gamma$) reaction
\cite{yokoi83} with the branching factor $^{\mbox{\small{Ta}}}\!f^m_{n}$.
For the relative abundance of $^{180}$Ta$^{m}$ one then obtains:
\begin{equation}\label{hauefigkeitfuer179hf}
\left[\frac{N_s(^{180}\mbox{Ta}^{m})}{N_s(^{178}\mbox{Hf})}\right]_2 =
\frac{\langle \sigma \rangle_{^{178}\mbox{\small{Hf+n}}}}{\langle \sigma 
\rangle_{^{180}\mbox{\small{Ta}}^{m}\mbox{\small{+n}}}}\cdot 
^{\mbox{\small{Ta}}}\!\!f^m_{n} \cdot f_{180} \ ,
\end{equation}
where $f_{180}$ is the branching factor for neutron captures at $^{178}$Hf 
leading to $^{180}$Ta,
\begin{equation}\label{f180hier}
f_{180}= \left\{\frac{\lambda(^{179}\mbox{Hf+n})}
{\lambda(^{179}\mbox{Hf})_{\beta^-}}
\cdot \left[\frac{\lambda(^{179}\mbox{Ta})_{EC}}{\lambda(^{179}\mbox{Ta+n})}
+1\right]+1 \right\}^{-1}
\end{equation}
and $\lambda$ are transition rates.
For $kT=30$ keV, $\langle \sigma \rangle_{^{178}\mbox{\small{Hf+n}}} = 310$ mb
\cite{beer:82s}, $^{\mbox{\small{Ta}}}\!f^m_{n} \approx (4.3\pm 0.8)\%$ 
\cite{nemeth:92}. The other parameters are taken from 
\cite{takahashi:87} (an error of $\pm 30\%$ assumed) and interpolated
for the $n_e$ and $T$.
For $kT=30$~keV this s-process branching yields $(190\pm 40) \%$ of 
the $^{180}$Ta$^{m}$ solar abundance.

It should be noted that all $^{180}$W can be produced in the s-process via 
the decay of the $^{180}$Ta ground-state.  For the relative abundance of 
$^{180}$W one can write
\begin{equation}
\label{180W}
\frac{N_s(^{180}\mbox{W})}{N_s(^{178}\mbox{Hf})}=
\frac{\langle \sigma \rangle_{^{178}\mbox{\small{Hf+n}}}}{\langle \sigma 
\rangle_{^{180}\mbox{\small{W+n}}}}\cdot 
\left( 1-^{\mbox{\small{Ta}}}\!\!f^m_{n}\right) \cdot 
f_{180} \cdot f_{\beta^-} \ ,
\end{equation}
where $f_{\beta^-}$ is the branching factor for the $^{180}$Ta ground-state
decay to $^{180}$W (numerical values taken from \cite{nemeth:92,takahashi:87}).
For $kT = 30$ keV this process can account for $(95 \pm 36)\%$ of the 
$^{180}$W solar abundance.
\vspace*{-0.2 cm}
\item In the s-process during the He shell burning in the AGB phase of 1.5 to 
3$M_{\odot}$ mass stars about 85\% of $^{180}$Ta$^{m}$ can be
produced \cite{Wisshak}.
\vspace*{-0.2 cm}
\item The p-process in higly evolved  massive stars: 
During the presupernova phase under temperatures
$T > 10^9$ K, thermal photons can induce  the reaction 
$^{181}$Ta($\gamma$,n) populating $^{180}$Ta$^{m}$ \cite{arnould:76}.
\vspace*{-0.2 cm}
\item $^{180}$Ta$^{m}$ production in the cosmic radiation:
Protons from the low-energy component of the galactic cosmic radiation 
can produce $^{180}$Ta$^{m}$ via the (p,yp~xn) reaction on s-process
or r-process nuclei in the interstellar medium \cite{haine:76}.  
\end{enumerate}

\begin{center}
\begin{figure}
\psfig{file=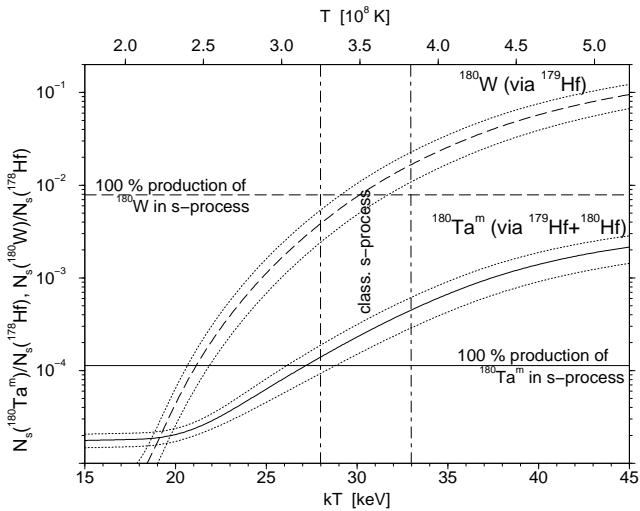,width=8.5cm}
\caption{\label{179Tavergl}
The total s-process abundance of $^{180}$Ta$^{m}$ (solid line) and 
$^{180}$W (long-dashed line) calculated relative to the $^{178}$Hf 
abundance as a function of the thermal energy $kT$ or temperature $T$.
Error bands are depicted by dotted lines. The s-process temperature window 
\protect\cite{Wisshak:95} is marked by dot-dashed lines.}
\end{figure}
\end{center}
\vspace*{-0.9 cm}

The total relative s-process abundance of $^{180}$Ta$^{m}$ can be calculated 
from:
\begin{eqnarray}\label{180sproc}
\frac{N_s(^{180}\mbox{Ta}^{m})}{N_s(^{178}\mbox{Hf})} & = &
\frac{N_s(^{179}\mbox{Hf})}{N_s(^{178}\mbox{Hf})} \cdot
\left[\frac{N_s(^{180}\mbox{Ta}^{m})}{N_s(^{179}\mbox{Hf})}\right]_1 
\nonumber \\
& + & \left[\frac{N_s(^{180}\mbox{Ta}^{m})}{N_s(^{178}\mbox{Hf})}\right]_2
\end{eqnarray} 
From Fig.~\ref{179Tavergl} one can see that for $kT=30$ keV an amount of 
$^{180}$Ta$^{m}$ $(2.0\pm0.5)$ times larger than the observed one can be 
produced in the classical s-process.

\begin{center}
\begin{figure}
\psfig{file=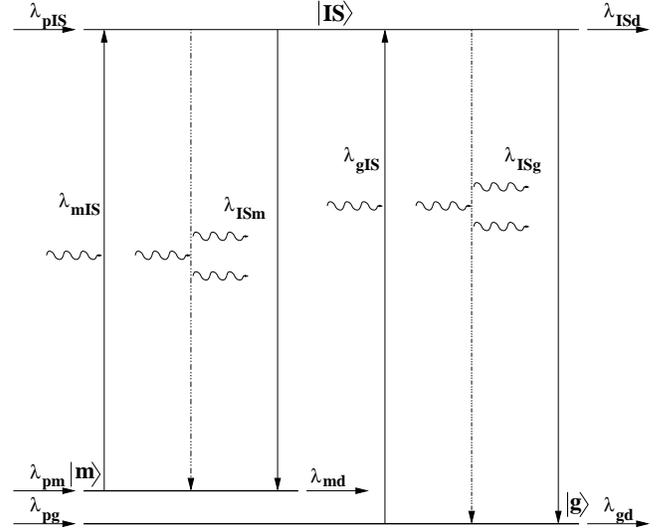,width=8.5cm}
\smallskip
\caption{\label{dreiniveau}Internal and external population and depopulation
possibilities in the three-level system described by means of transition rates
$\lambda$ \protect\cite{ward:80}.}
\end{figure}
\end{center}
\vspace*{-0.9 cm}

This overproduction is reduced by intermediate states (IS) in 
$^{180}$Ta: In 
the s-process site thermal photons may excite higher-lying levels which then 
decay back either to the $1^{+}$ ground state or to the $9^{-}$ isomer. 
To find these levels Belic {\it et al.} \cite{Belnew,Bel} used the Stuttgart 
Dynamitron facility with both enriched ($5.6\%$) and natural Ta targets. 
Irradiations were performed for bremsstrahlung endpoint energies 
$E_0 = 0.8$--$3.1$ MeV. Depopulation of the isomer was observed down to 
$E_0 \approx 1.01$ MeV. This means that the lowest IS may have an 
excitation energy $E_{IS} = 1.085$ MeV (above the ground state).
Assuming that $E_{IS}$ is the excitation energy of the lowest IS,
the experimental total integrated depopulation
cross section $I_D$ turns out to be $(5.7\pm 1.2)$ eV fm$^2$. Then the 
effective lifetime 
of the IS for the depopulation into the ground state 
$\tau_{\mbox{\small{eff}}}$ is roughly equal to $6 \cdot 10^{-11}$ s.

The IS $|\mbox{IS}\rangle$ decays via a $\gamma$-cascade,
and lifetimes and energies of the decay states $|\mbox{k}\rangle$ are not 
known. Klay \cite{klay:90} showed that multi-step transitions from 
$|\mbox{IS}\rangle$ to the ground-state $|\mbox{g}\rangle$ can be substituted 
by a direct transition from $|\mbox{IS}\rangle$ to $|\mbox{g}\rangle$ if
the states $|\mbox{k}\rangle$ posses short enough lifetimes $\tau_k$, i.e.
\begin{equation}\label{time}
\tau_k \ll \tau_{\mbox{\small{eff}}} \cdot \frac{2I_k+1}{2I_{IS}+1}\cdot 
\exp{({|E_{IS}-E_k|/kT})}
\end{equation}
where $E_{IS}$ and $I_{IS}$, $E_k$ and $I_k$ are the energies and spins of the 
IS and the states $|\mbox{k}\rangle$, respectively. 
It can be shown that the condition (\ref{time}) is fulfilled in 
$^{180}$Ta and transitions from the isomer $|\mbox{m}\rangle$ to the
ground-state $|\mbox{g}\rangle$ can be studied in the three-level system (see
Fig.~\ref{dreiniveau}).	

The population of the three-level system can be described by the following 
coupled inhomogeneous linear differential equations:
\begin{eqnarray}\label{diffglsys}
\frac{dN_{m}}{dt} & = & \lambda_{pm}\!~N_s-(\lambda_{mIS}+
\lambda_{md})\!~N_m+\lambda_{ISm}\!~N_{IS} \nonumber \\
\frac{dN_{IS}}{dt} & = & \lambda_{pIS}\!~N_s+
\lambda_{gIS}\!~N_g+\lambda_{mIS}\!~N_m \nonumber \\
 & - &(\lambda_{ISg}+
\lambda_{ISm}+\lambda_{ISd})\!~N_{IS} \\
\frac{dN_{g}}{dt} & = & \lambda_{pg}\!~N_s-(\lambda_{gIS}+
\lambda_{gd})\!~N_g+\lambda_{ISg}\!~N_{IS} \nonumber \ ,
\end{eqnarray}
where $\lambda_{px}$ ($\lambda_{xd}$) are the population (depopulation) 
transition rates of the state $|x\rangle$, $\lambda_{xy}$ are the transition
rates between the states $|x\rangle$ and $|y\rangle$, $N_x$ is the number
of nuclei in the state $|x\rangle$, $N_x (t=0)=0$, and $N_s$ is a constant 
number of seed nuclei ($^{178}$Hf in our case).
 
In previous analyses, e.g. \cite{Bel,Schumann}, the effect of an IS on the 
survival of $^{180}$Ta$^{m}$ in the presence of a stellar photon bath was 
calculated by solving coupled differential equations for the three-level
system isomer $\leftrightarrow$ IS $\leftrightarrow$ ground-state without
taking into account the population of the three levels due to the s-process
simultaneously, i.e. $\lambda_{pm}=\lambda_{pIS}=\lambda_{pg}=0$ was assumed.

\begin{center}
\begin{figure}
\psfig{file=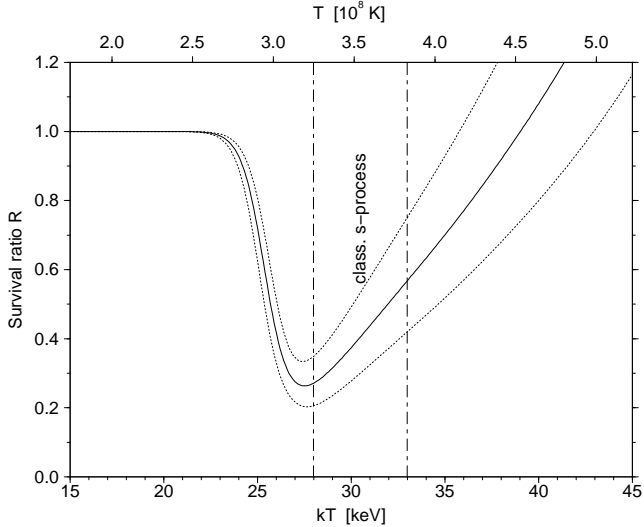,width=8.5cm}
\caption{\label{n0_von_kT}Survival ratio $R$ of $^{180}$Ta$^{m}$ as a function
of $kT$ or $T$ for the IS at $E_{IS}=1.085$ MeV. Error bands are marked by 
dotted lines, the s-process temperature window by dot-dashed lines.}
\end{figure}
\end{center}
\vspace*{-0.9 cm}

For $t \rightarrow \infty$ the solution of (\ref{diffglsys}) approaches the
equilibrium solution. Since the maximum equilibrium relaxation time 
$\tau_{max}$ ($\tau_{max}<100$ days for $kT$ between 15 keV and 45 keV) is 
smaller than the s-process duration ($> 1$ year) the exact solution of 
(\ref{diffglsys}) can be approximated by the equilibrium solution. 
In the case of existing IS one can define the survival ratio $R$ of 
$^{180}$Ta$^{m}$ (depicted in Fig.~\ref{n0_von_kT}):
\begin{equation}
R\equiv \frac{N_m(t \rightarrow \infty )}{N_s(^{178}\mbox{Hf})} \Big/ 
\frac{N_s(^{180}\mbox{Ta}^{m})}{N_s(^{178}\mbox{Hf})}
\end{equation} 
where $N_s(^{178}\mbox{Hf})=N_s$ in (\ref{diffglsys}), 
$N_s(^{180}\mbox{Ta}^{m})/N_s(^{178}\mbox{Hf})$ represents the
relative abundance for no IS and is obtained from 
(\ref{180sproc}). The survival ratio $R$ is larger then 0.2 in the 
whole temperature interval. For $kT\leq 21$~keV the coupling between the 
isomer and the ground state via the IS is negligible. Note that $R>1$ for 
$kT \stackrel{>}{\sim} 40$~keV.

It can be shown that for $kT=30$ keV only dominant transition
rates $\lambda $ must be taken into account and that the equilibrium solution 
can be approximated by
\begin{equation}\label{approximation}
\frac{N_m(t \rightarrow \infty )}{N_s(^{178}\mbox{Hf})} =
\left[\frac{N_s(^{180}\mbox{Ta}^{m})}{N_s(^{178}\mbox{Hf})}\right]_{IS}
\approx {\Large \frac{\lambda_{pm}+\lambda_{pg}+\lambda_{pIS}}
{\lambda_{gd}/P+\lambda_{md}}}
\end{equation}
where
\begin{equation}
P=\frac{2I_m+1}{2I_g+1}\cdot \exp{(-75.3~\mbox{keV}/kT)} \approx 0.5 \ ,
\end{equation}
\begin{equation}
\lambda_{md} = \lambda\mbox{(}^{180}\mbox{Ta}^{m}\mbox{+n)}= n_n \cdot
\left(\frac{2kT}{m_n}\right)^{1/2} \langle \sigma 
\rangle_{^{180}\mbox{\small{Ta}}^{m} \mbox{\small{+n}}} \ ,
\label{lamsig}
\end{equation}
$\lambda_{md} \approx 10^{-7}$ s$^{-1}$, $m_n$ is the neutron mass, 
$\lambda_{gd} = \lambda$($^{180}$Ta)$_{\beta^-}+
\lambda$($^{180}$Ta)$_{EC} = 4.2 \cdot 10^{-6}$ s$^{-1}$ \cite{takahashi:87},
\begin{equation}
\lambda_{pm}+\lambda_{pg}+\lambda_{pIS} \approx 
\lambda\mbox{(}^{178}\mbox{Hf+n)} \cdot \left( f_{180} + 
^{\mbox{\small{Hf}}}\!f^m_{n} \cdot 
f^{m}_{\beta ^-} \right) \ ,
\label{production}
\end{equation}
where $\lambda$($^{178}$Hf+n) can be calculated from 
$\langle \sigma \rangle_{^{178}\mbox{\small{Hf+n}}}$. 
The two terms in (\ref{production}) originate in the two s-process 
branchings via $^{179}$Hf and $^{180}$Hf$^m$ that can produce $^{180}$Ta.
As a consequence of (\ref{approximation}) the approximate solution does not 
depend on properties of the IS like its spin, energy, lifetime and transition 
rates to isomer and ground state.

\begin{center}
\begin{figure}
\psfig{file=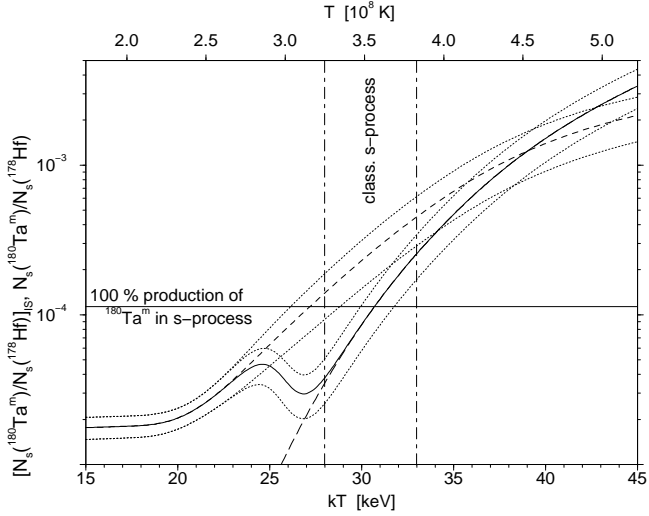,width=8.5cm}
\caption{\label{180TaVmZ}
Comparison of the total s-process abundances of $^{180}$Ta$^{m}$ calculated 
relative to the $^{178}$Hf abundance as a function of the thermal energy 
$kT$ or temperature $T$: for the IS at $1.085$ MeV the equilibrium solution 
of \protect(\ref{diffglsys}) is denoted by a bold solid line, and the 
approximate solution \protect(\ref{approximation}) by a bold long-dashed 
line, and the total abundance for no IS from \protect(\ref{180sproc}) by 
a bold dashed line. Dotted lines correspond to error bands, dot-dashed lines
to the s-process temperature window.}
\end{figure}
\end{center}

In Fig.~\ref{180TaVmZ} the relative abundance of $^{180}$Ta$^{m}$ as a function
of $kT$ or $T$ is depicted. The equilibrium solution of (\ref{diffglsys}) 
with the IS at $1.085$~MeV is compared to the case of no IS 
(cf.~Fig.~\ref{179Tavergl}).
Note that for $kT \geq 29$~keV the approximate (\ref{approximation})
and the equilibrium solution
are indistinguishable. As can be seen in Fig.~\ref{180TaVmZ} exactly 100\% 
of the solar $^{180}$Ta$^{m}$ abundance can be reproduced in the middle of the
s-process temperature window. Other IS that may be found in the future will
not change this result (cf. a possible IS below 737 keV \cite{lakosi}).

If the IS at 1.085 MeV in $^{180}$Ta is taken into account the relative 
abundance of $^{180}$W (\ref{180W}) must be corrected:
\begin{equation}
\left[\frac{N_s(^{180}\mbox{W})}{N_s(^{178}\mbox{Hf})}\right]_{IS} \approx
\frac{\langle \sigma \rangle_{^{178}\mbox{\small{Hf+n}}}}{\langle \sigma 
\rangle_{^{180}\mbox{\small{W+n}}}}\cdot 
\left( 1-^{\mbox{\small{Ta}}}\!\!f^m_{n} R \right) 
\cdot f_{180} \cdot f_{\beta^-} \ .
\end{equation}
Since the survival ratio $R \le 1$ for the s-process temperature window
(see Fig.~\ref{n0_von_kT}) and $^{\mbox{\small{Ta}}}\!f^m_{n} = 4.3\%$ 
the maximum change of the relative abundance of $^{180}$W due to the IS 
represents about $4\%$ and can be neglected. 

\vspace*{-0.9 cm}

We have shown that $^{180}$Ta$^{m}$ could be produced in the classical 
s-process dominantly via the branching in $^{179}$Hf. The production of 
$^{180}$Ta$^{m}$ is then connected to the production of $^{180}$W. 
For the temperature window $kT = (28 - 33)$ keV we obtain 
$(50\pm 20) \% - (220 \pm 80) \%$ of $^{180}$W produced in the s-process 
via $^{180}$Ta. Assuming sufficiently low temperatures in the star evolution 
phases following the s-process and the IS at $E_{IS} \le 1.085$ MeV as observed
in the Stuttgart photoactivation experiment \cite{Bel,Belnew}, the s-process 
production of $^{180}$Ta$^{m}$ ranges from $(30\pm 10)\%$ to $(230\pm 80)\%$ 
of its observed solar abundance for the same temperature window. 
For the middle of the temperature window ($kT=30.5$~keV) we obtain
$(90\pm 30)\%$ of $^{180}$Ta$^{m}$ (see Fig.~\ref{180TaVmZ}) and 
$(110\pm 40)\%$ of $^{180}$W. It should be noted that $^{180}$W is often
listed among p-process nuclei \cite{rayet:95}. An upper limit for its 
production in the s-process provided by an improved model of the 
p-process would have direct consequences for the 
$^{180}$Ta$^{m}$ production in the classical s-process via the branchings in 
$^{179}$Hf and $^{180}$Hf. 

\vspace*{-0.6 cm}
\acknowledgements
\vspace*{-0.4 cm}
The authors are grateful to Dr. F.\ K{\"a}ppeler and D.\ Brandmaier 
for valuable discussions.

\end{document}